# Spectroscopic Evidence for Interfacial Charge Separation and Recombination in Graphene-$MoS_2$ Vertical Heterostructures

Yuqing Zou[1, §], Zeyu Zhang[2, §], Chunwei Wang[2,3], Yifan Cheng[1], Chen Wang[1], Kaiwen Sun[1], Wenjie Zhang[1], Peng Suo[1], Xian Lin[1], Hong Ma[4], Yuxin Leng[2], Weimin Liu[3,*], Juan Du[2,*], Guohong Ma[1,*]

[1]Department of Physics, Shanghai University, Shanghai 200444, China

[2]School of Physics and Optoelectronic Engineering, Hangzhou Institute for Advanced Study, University of Chinese Academy of Sciences, Hangzhou,310024, China

[3]School of Physical Science and Technology, ShanghaiTech University, Shanghai 201210, China

[4]School of Physics and Electronics, Shandong Normal University, Jinan 250014, China

[§]These authors contributed Equally

[*] Email address: liuwm@shanghaitech.edu.cn (W. M. Liu), dujuan@mail.siom.ac.cn (J. Du), and ghma@staff.shu.edu.cn (G. H. Ma)

**ABSTRACT:**

Vertical van der Waals (vdW) heterostructures consisting of graphene (Gr) and transition metal dichalcogenides (TMDs) have created a fascinating platform for exploring optical and electronic properties in the two-dimensional limit. Previous study has revealed the ultrafast formation of interfacial excitons and the exciton dynamics in the Gr/$MoS_2$ heterostructure. However, a fully understanding of interfacial charge separation and the subsequent dynamics in graphene-based heterostructures remains elusive. Here, we investigate the carrier dynamics of Gr-$MoS_2$ (including Gr/$MoS_2$ and $MoS_2$/Gr stacking sequences) heterostructures under different photoexcitation energies and stacking sequences by comprehensive ultrafast means, including time-resolved terahertz spectroscopy (TRTS), terahertz emission spectroscopy (TES) and transient absorption spectroscopy (TAS). We demonstrate that the Gr/$MoS_2$ heterostructure generates hot electron injection from graphene into the $MoS_2$ layer with photoexcitation of sub-A-exciton of $MoS_2$, while the interfacial charge separation in the $MoS_2$/Gr could be partially blocked by the electric field of substrate. Charge transfer (CT) occurs in same directions for the Gr-$MoS_2$ heterostructures with opposite stacking order, resulting in the opposite orientations of the interfacial photocurrent, as directly demonstrated by the terahertz (THz) emission. Moreover, we demonstrate that the recombination time of interfacial charges after CT is on a timescale of 18 ps to 1 ns, depending on the density of defect states in $MoS_2$ layer. This work provides a comprehensive and unambiguous picture of the interfacial charge dynamics of graphene-based heterostructures, which is essential for developing Gr/TMDs based optoelectronic devices.

## I. INTRODUCTION:

Atomically thin two-dimensional materials, including monolayer (ML) graphene (Gr) and transition metal dichalcogenides (TMDs), have attracted extensive attention in the field of optoelectronic applications. ML Gr has been proposed as a promising optoelectronic material with zero band gap, high carrier mobility, wide band optical absorption, *etc.*[1-3] However, the photo-electric response in Gr is limited by the fast carrier lifetime and low absorption rate in the visible spectrum.[4] Complementarily, TMDs, such as ML $MoS_2$, can perfectly compensate for by their peculiar properties such as spin-valley coupling,[5,6] visible band-gap energy,[7-9] and strong optical absorption and nonlinear optical response.[10-12] In fact, van der Waals (vdW) heterostructures formed by vertical stacking of graphene and TMDs (Gr-TMDs) break the limitation of lattice mismatch and benefit from the advantages of each layer,[13-17] which greatly extends the feasibility and compatibility for developing novel heterostructures. Furthermore, vdW heterostructures can be controlled by the number of atomic layers, stacking order, twist angle, and external electric field to tailor specific properties for target applications, showing great promise for optical,[18] optoelectronics,[19,20] and electrochemical applications,[21-23]*etc.*

Understanding photoinduced interfacial charge separation processes in heterostructures is crucial for optoelectronic applications, and numerous recent studies have investigated charge and/or energy transfer in Gr/TMDs heterostructures *via* ultrafast spectroscopy.[13, 24-36] It has been reported that the charge transfer (CT) process of Gr/TMDs heterostructures interface is very fast, with carrier transfer completed within ~20 fs.[28] Currently, interfacial photothermionic transfer upon sub-band gap photoexcitation has been widely accepted,[24-29] where TMDs act as a channel to collect thermal charge carriers from graphene layer before the cooling process occurs, resulting the transferred electrons and holes reside in different atomic layers. However, the recombination time of the interfacial charge remains elusive. For instance, Yuan *et al.* reported a rapid charge-separation lifetime of ~1 ps,[13] which is extremely similar to the results obtained by Aeschlimann *et al.* using time- and angle-resolved photoemission spectra.[30] In contrast, Fu *et al.* reported long-lived charge separation states of ~1 ns.[26] While our previous study of the Gr/$MoS_2$ heterostructure with sub-band gap excitation demonstrates a typical interfacial exciton state of ~18 ps.[25] In addition, factors affecting interfacial CT in heterostructures have recently been investigated by a series of studies. Liu *et al.* reported that an external electric field can efficiently modulate the carrier dynamics at the interface of the Gr/$WS_2$ heterostructure,[35] and Fu *et al.* further controlled the charge-separated state and the direction of the photogating field by electrically tuning the defect occupancy.[36] Luo *et al.* experimentally verified the tuning effect of the twist angle on the photoresponse in the $MoS_2$/Gr heterostructure using ultrafast electron diffraction.[31] However, as far as we know, the tuning effect of stacking order on the CT in Gr-TMDs heterostructures has rarely been reported. Moreover, the interfacial charge recombination time, and in particular the dynamical behavior of excitons, remains to be elusive.

In this report, in order to portray the comprehensive and unambiguous picture of interfacial charge separation and excitonic behavior of graphene-based heterojunctions, we investigate the ultrafast carrier dynamics of Gr-$MoS_2$ (including Gr/$MoS_2$ and $MoS_2$/Gr stacking sequences) heterostructures under different photoexcitation energies and stacking sequences *via* time-resolved terahertz spectroscopy (TRTS), terahertz emission spectroscopy (TES) and transient absorption spectroscopy (TAS). The spectroscopic results reveal that when the photoexcitation energy is below the A-exciton resonance of $MoS_2$, the surface-directed substrate electric field facilitates the hot electrons injection from graphene into the $MoS_2$ layer in the Gr/$MoS_2$ heterostructure, by contrast, the electrons injection in the $MoS_2$/Gr heterostructure is suppressed significantly by the substrate electric field. Furthermore, with above A-exciton photoexcitation, the experimental results demonstrate that the interfacial CT directions is the same for the two heterostructures independently of the stacking order, namely, the direct transfer of $MoS_2$ valence band holes to the graphene layer occurs in both Gr/$MoS_2$ and $MoS_2$/Gr heterostructures, leading to the opposite orientations of the interfacial photocurrent, this is directly demonstrated by THz emission measurements.

Remarkably, comparing the hole lifetime in graphene layer obtained from TRTS with the electron lifetime in $MoS_2$ *via* TAS, our data support that the interfacial recombination of separated charges occurs on a timescale of 18 ps to 1 ns following CT, and the relaxation time is found to strongly depend on the density of defect states in the $MoS_2$ layer. This study provides substantial evidence for interfacial CT and excitonic behavior in Gr-TMDs heterostructures, as well as a dynamic picture for understanding complex multibody interactions, which is expected to facilitate the application of graphene-based optoelectronic devices.

## II. EXPERIMENTAL METHODS

**Samples Preparation.** Monolayers of graphene and $MoS_2$, as well as Gr-$MoS_2$ heterostructures on silica substrates are obtained commercially (provided by Six Carbon Technology, Shenzhen, China) and prepared by CVD method. The samples are characterized by UV-vis and Raman spectroscopy to test the high quality of the monolayer properties of the thin film samples.

**TES and TRTS.** The THz spectral system is pumped by a femtosecond laser (based on regenerative amplification and mod-locked Ti: Sapphire) with a central wavelength of 780 nm (1.6 eV), a duration of 120 fs, and a repetition rate of 1 kHz. In TRTS measurements, the laser pulses delivered from an amplifier are divided into three beams: the first two are used to generate and detect terahertz signal, configuring a terahertz time-domain spectral system to measure the static conductivity of the sample, where the THz emitter and detector are based on a pair of (110)-oriented ZnTe crystals. The third beam is a pump beam, which is used to excite the sample, and the pump beam is set to propagate collinearly with the THz beam. Optical pumps for other wavelengths are output by optical parameter amplifiers (TOPS prime). In TES, a fs pump beam is focused directly onto the sample and the THz pulse emitted by the sample is focused onto the ZnTe detection crystal by the off-axis parabolic mirrors assemblies. The optical pump and THz detection pulses are collinearly polarized and have spot sizes of 6.5 and 2.0 mm on the sample surface, respectively.

**TAS.** The TAS system is a commercial TA spectrometer (HELIOS FIRE, Ultrafast

System) driven by a Ti: sapphire laser (Coherent, Astrella). The laser outputs optical pulses centered at 800 nm with a pulse duration of 35 fs and repetition rate of 1 kHz, which was divided into two beams: one was guided into an optical parametric amplifier to produce 780 nm excitation light, and the other beam was focused on a sapphire slice to generate white light supercontinuum with the pulse duration of 150 fs, which serves as a probe beam in the ultrafast system. By controlling the time delay of the pump-probe with an electrically translation platform, not only the absorption values at different detection wavelengths under a specific time, but also the dynamics process of the sample at a specific detection wavelength can be obtained. The change of absorbance (ΔA) induced by the pump optical pulse reflects the photoinduced bleaching ($\Delta A<0$) and/or photoinduced absorption ($\Delta A>0$). During the measurement, samples were placed on a holder and moved within a 1×1 $mm^2$ plane perpendicular to the probe light propagation direction to avoid heat buildup. All the measurements were performed at room temperature.

## III. RESULTS AND DISCUSSION:

**Characterization of samples and experimental methods.** The graphene and $MoS_2$ monolayers used in this study were grown by CVD method on a 1 mm-thick silica substrate (provided by Sixcarbon Tech, Shenzhen, China). The bilayer heterostructures with opposite stacking sequence, *i.e.* Gr/$MoS_2$/$SiO_2$ and $MoS_2$/Gr/$SiO_2$, were fabricated by CVD and liquid transfer method on a silica substrate. **Figure 1**a illustrates schematically the two-type heterostructures. The characteristic peaks of ML $MoS_2$ with the ultraviolet-visible absorption spectra are located at ~1.89 and ~2.04 eV, shown in **Figure S1a in the Supporting Information (SI)**, corresponding to the recognized A- and B-exciton transitions, respectively.[37,38] Upon contact with graphene, the A- and B-exciton resonances of the Gr/$MoS_2$ and $MoS_2$/Gr heterostructures are red-shifted and blue-shifted, respectively, both to amount of about 60 meV. We attribute the stacking-order related spectra shift to the interlayer coupling effect between Gr and $MoS_2$ layers, which modifies the $MoS_2$ layer energy gap and causes a slight shift in the exciton absorption peak. In addition, the characteristic Raman spectra of the samples are shown in **Figure S1b in the SI**. For Gr/$MoS_2$ and $MoS_2$/Gr heterostructures, the characteristic peaks observed around ~ 1587 and ~ 2675 $cm^{-1}$ are the G- and 2D bands of the graphene layer, respectively, indicating the high quality of the graphene monolayer in the heterostructures,[39,40] and the Raman peaks observed near ~384 and ~404 $cm^{-1}$ are typical $E_{1g}^{2}$ and $A_{1g}$ modes of $MoS_2$ monolayer, respectively.[8,19] Notably, the Raman peak shift between the heterostructures and ML $MoS_2$ indicates that the $SiO_2$ substrate introduces an effective internal electric field,[19,41,42] which points outward towards the surface. It is also noted that the influence of substrate electric field on $MoS_2$ is more pronounced in the Gr/$MoS_2$ than that in $MoS_2$/Gr, which is probably due to the screening effect of graphene layer on the electric field in $MoS_2$/Gr heterostructure, as shown schematically in **Figure 1**a (**see Section S1 in the SI for more details**).

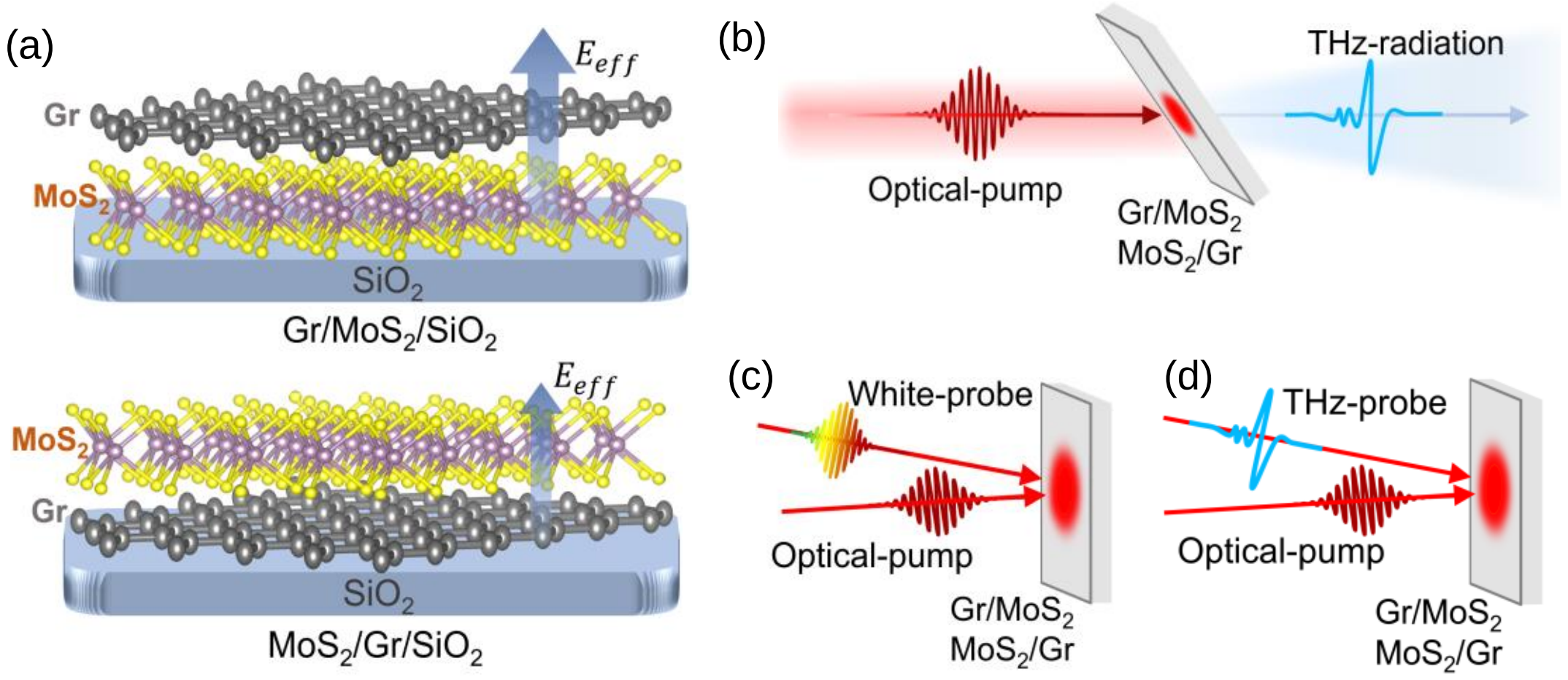

Figure 1. Schematic diagrams of Gr-$MoS_2$ heterostructures and experimental methods. (a) Schematic of the vertically stacked Gr/$MoS_2$ and $MoS_2$/Gr heterostructures, with the blue arrows indicating the effective electric fields introduced by the $SiO_2$ substrate. (b-d) Spectroscopic diagrams of TES, TAS and TRTS, respectively.

TES measurement is shown in **Figure 1**b. It is expected that the interfacial carrier transport of Gr-$MoS_2$ heterostructures can generate ultrafast directional photocurrents along the stacking direction, resulting in THz radiation, and thus TES is used to probe the interfacial CT processes in the heterostructures.[44,45] In addition, TAS in the visible region is an excellent spectroscopic means for studying the excited state carrier dynamics of $MoS_2$. As well, TRTS has proven to be an essential tool for probing the dynamics of free carriers due to its sensitivity to the conductivity change of carrier excitations, as well as the relaxation and recombination processes. **Figures 1**c and **1**d show the measurement schematics for TAS and TRTS, respectively, see the Experimental Methods section for details on the specific experimental setup. Therefore, a combination of several ultrafast spectroscopic methods described above are employed to obtain comprehensive information on the non-equilibrium carrier dynamics at the interface of Gr-$MoS_2$ heterostructures. Here, we tune the pump photoexcitation to selectively excite only the graphene layer ($h\nu$<1.89 eV, namely below A-exciton excitation) or both layers together ( $h\nu$>1.89 eV, on resonance or above A-excitation) in heterostructures. The photoconductivity information $\sim\Delta\sigma$ of the samples can be obtained by monitoring the THz electric field variation $\sim\Delta E$ $(\Delta E = E_{pump} - E_0)$ before and after optical pumping. The measurement principle applies to the proportional relationship between $\Delta\sigma$ and $\Delta E$, as described in equation 1,[25,43]

$$\Delta\sigma = -\frac{(1+n)}{Z_0}\frac{\Delta E(t)}{E_0(t)} \tag{1}$$

where $n$=1.95 is refractive index of $SiO_2$ substrate, $Z_0$=377 is free-space impedance, and $E_{pump}$ and $E_0$ denote the transmitted THz electric field with and without optical excitation, respectively.

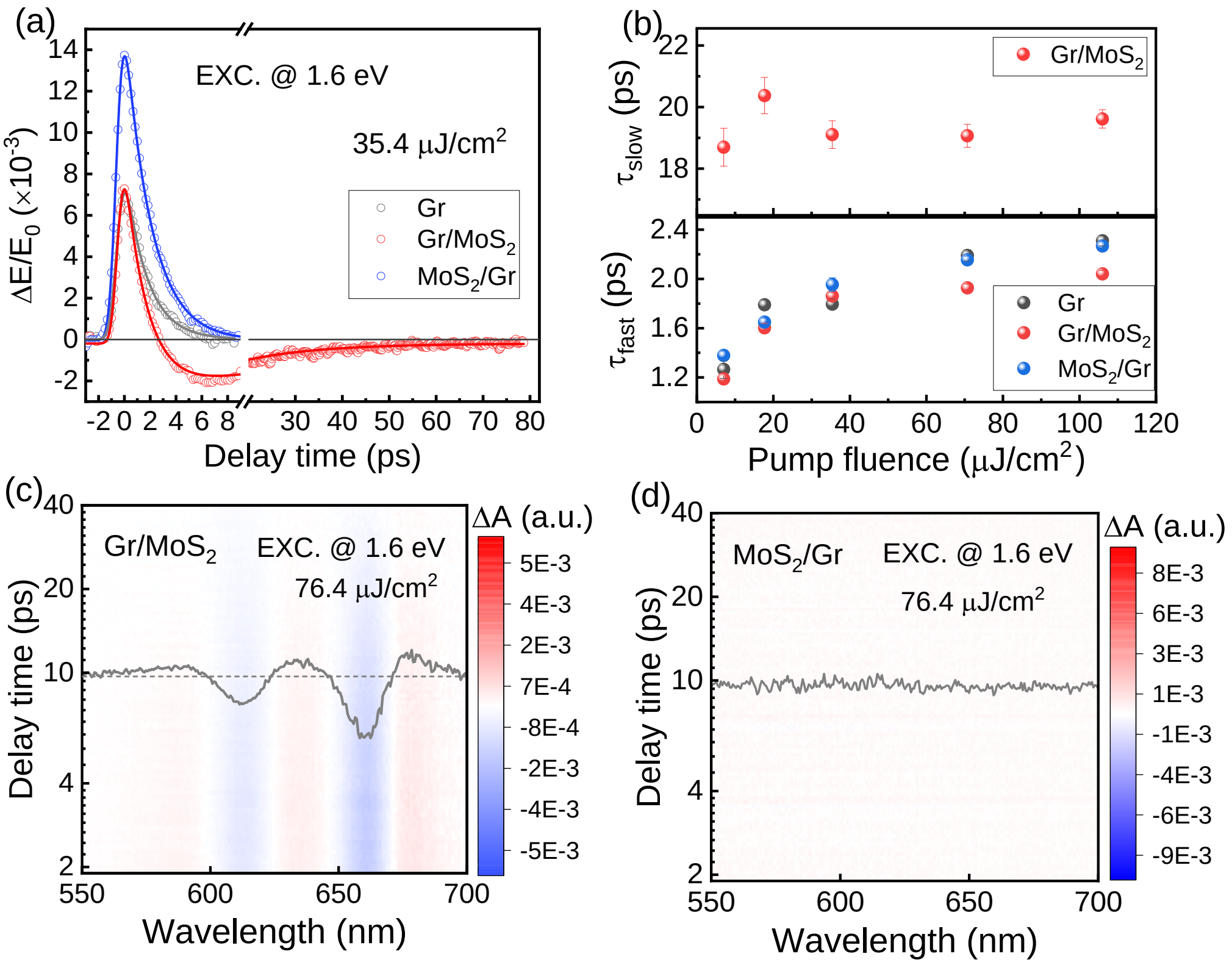

Figure 2. Interfacial CT of Gr-$MoS_2$ heterostructures under 1.6 eV (780 nm) excitation. (a) Transient THz transmittance of Gr (gray), Gr/$MoS_2$ (red) and $MoS_2$/Gr (blue) heterostructures with a pump fluence of 35.4 $\mu$J cm$^{-2}$. The solid lines are the fitted curves with equation 2 in the main text. (b) Relaxation time of THz photoconductivity of Gr, $MoS_2$/Gr (by single exponential fit) and Gr/$MoS_2$ heterostructures (by biexponential fit) with respect to pump fluence. (c and d) 2D plots with TAS of Gr/$MoS_2$ and $MoS_2$/Gr heterostructures with fixed pump fluence ~76.4 $\mu$J cm$^{-2}$, respectively.

**Photocarrier dynamics in Gr-$MoS_2$ heterostructures with sub-A-exciton photoexcitation.** The THz photoconductivity of ML Gr, ML $MoS_2$ and Gr-$MoS_2$ heterostructures was tested by TRTS with 1.6 eV (780 nm) photoexcitation, and the pump fluence was fixed at 35.4 $\mu$J cm$^{-2}$. Since the photoexcitation energy (1.6 eV) is lower than the A-exciton resonance (1.89 eV) of $MoS_2$, no signal is detected in ML $MoS_2$. As shown in **Figure 2**a, the ML Gr exhibits a negative photoconductivity, while the photoconductivity polarity of Gr/$MoS_2$ varies with delay time. Our previous work has demonstrated that the positive photoconductivity of Gr/$MoS_2$ is due to interfacial CT, where the extraction of graphene thermal electrons leads to a decrease in the Fermi level and an increase in the conductivity.[25] Interestingly, the reverse-stacked $MoS_2$/Gr heterostructure exhibits the same negative photoconductivity as ML Gr, *i.e.*, no significant interfacial CT is observed in $MoS_2$/Gr. To further confirm our conjecture, the data in **Figure 2**a are well fitted by an exponential decay function convolved with the Gaussian function, which is depicted in equation 2[25], where $\tau$ and $A$ are relaxation time and corresponding amplitude, respectively, 2ω is the full width at half-maximum (fwhm) of the THz waveform, *erfc*(*x*)=1-*erf*(*x*) is the complementary error function. See **Figure S2** for the detailed fitting process.

$$\frac{\Delta E}{E_0}(t) = \sum_j A_j \times e^{\omega^2/\tau_j^2 - t/\tau_j} \times erfc\left(\omega/\tau_j - t/2\omega\right) \quad (2)$$

**Figure 2**b displays the fitting results, as expected, the Gr/$MoS_2$ heterostructure exhibits a typically slow interfacial exciton lifetime (18~20 ps) in addition to the fast thermal electron scattering lifetime (1~3 ps) of the graphene layer.[25] In contrast, the $MoS_2$/Gr heterostructure conforms to the single-exponential decay characteristic and maintains the same pump-dependent decay lifetime (1~3 ps) as ML Gr, as shown in the lower panel of **Figure 2**b, where no interfacial CT is detected. Furthermore, complementary measurements were performed using TAS for monolayers and heterostructures, again with the excitation pulse fixed at 1.6 eV. For ML Gr and $MoS_2$, as expected, no exciton bleaching signal was observed. **Figure 2**c and **2**d are the two-dimensional (2D) TA color maps for Gr/$MoS_2$ and $MoS_2$/Gr heterostructures, respectively, with a pump fluence of 76.4 $\mu$J $cm^{-2}$. In agreement with THz spectroscopic observations, the obvious bleaching signals around A- and B-exciton of $MoS_2$ in the Gr/$MoS_2$ heterostructure provides direct evidence for graphene electron injection into the $MoS_2$ layer.[25] However, no signal was detected in the $MoS_2$/Gr heterostructure, further confirming that CT across the heterostructure does not occur. In short, with sub-A-exciton photoexcitation, the hot electrons from graphene are injected into the $MoS_2$ layer in the Gr/$MoS_2$ heterostructure and promote the formation of interfacial excitons with lifetime of ~18 ps. By contrast, the interfacial CT in the $MoS_2$/Gr heterostructure is blocked and the hot electrons are still retained in the graphene layer for cooling. We attribute this to the modulation of the effective electric field introduced by the substrate in $MoS_2$/Gr heterostructure,[41,42] which prevents the transfer of thermal electrons across the interfacial barrier. The substrate electric field directed towards the surface (**Figure 1**a) promotes CT in the Gr/$MoS_2$ but suppresses the interfacial charge separation in the $MoS_2$/Gr heterostructure, as illustrated in **Figures 5**b and **5**c, respectively.

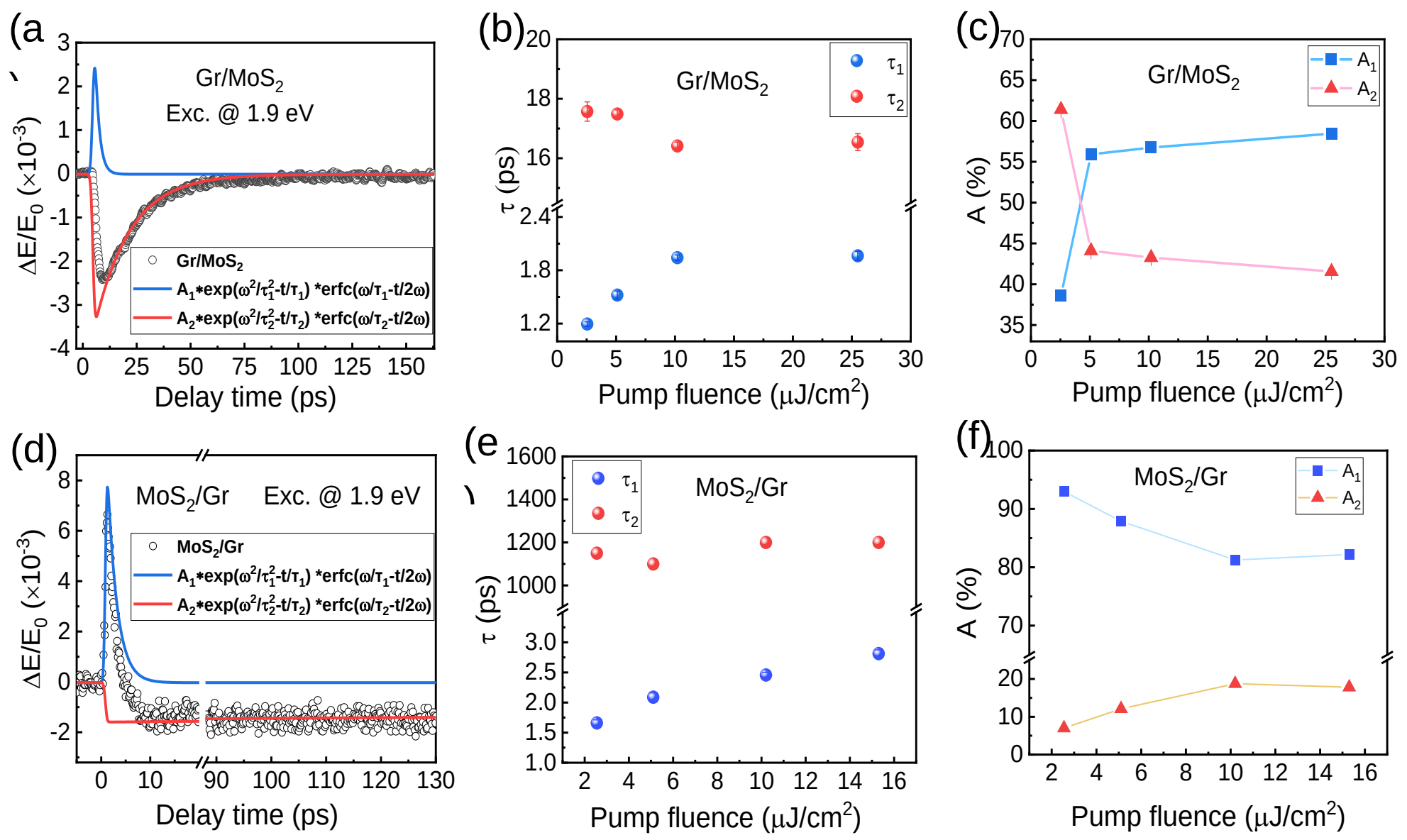

Figure 3. TRTS of Gr-$MoS_2$ heterostructures at 1.9 eV (650 nm) excitation. (a and f) Transient THz spectra (black circle) of the Gr/$MoS_2$ and $MoS_2$/Gr heterostructures with a pump fluence of 10.2 $\mu$J cm$^{-2}$, where the blue and red solid lines are the fitted curves for the biexponential decay function. (b and e) The fitted lifetimes as a function of the pump fluence for the Gr/$MoS_2$ and the $MoS_2$/Gr heterostructures, respectively. Comparison of the fitted amplitudes of the Gr/$MoS_2$ (c) and $MoS_2$/Gr heterostructures (f).

**Photocarrier dynamics in Gr-$MoS_2$ heterostructures with above A-exciton photoexcitation.** To further construct a complete charge transport model for Gr-$MoS_2$ heterostructures system, we performed the complementary spectral measurements of Gr/$MoS_2$ and $MoS_2$/Gr with photoexcitation of 1.9 eV (650 nm), near resonance with the A-exciton of $MoS_2$. In addition, we have tested the THz response of ML Gr and $MoS_2$ under the identical condition as a control. As expected, the ML Gr exhibits negative photoconductivity and follows a single exponential decay with a lifetime of 1~2 ps (**Figure S3d**). The ML $MoS_2$, however, shows no detectable signal due to the almost zero contribution of intramolecular bound excitons to the THz photoconductivity. **Figures 3**a and 3f show the transient THz transmission for Gr/$MoS_2$ and $MoS_2$/Gr heterostructures, respectively. Interestingly, there is a large difference in the photoconductivity between the two. Compared with the Gr/$MoS_2$ heterostructure, the negative photoconductivity of $MoS_2$/Gr is significant, followed by the positive photoconductivity with a long life (over 1 ns). For further analysis, we can still fit the experimental data with an exponential function. As shown in **Figure 3**a, the Gr/$MoS_2$ heterostructure perfectly fits the biexponential property, and the resulting decay lifetimes and weights obtained are displayed in **Figures 3**b and **3**c, respectively. It is clear that the fast lifetime $\tau_1$ (1~3 ps) comes from the negative photoconductivity of graphene layer. Moreover, the positive photoconductivity with a much stronger intensity shows a relaxation time of ~17 ps ($\tau_2$), which is extremely close to that with 1.6 eV excitation in **Figure 2**b. The graphene thin layers used in this experiment are all *p*-doped. Considering the large optical absorption of $MoS_2$ around A-exciton, so that the photogenerated holes in the top valence band of $MoS_2$ can be efficiently injected into the valance band of graphene, leading to a significant reduction of graphene Fermi energy level, which results in a pronounced positive THz photoconductivity in Gr/$MoS_2$ upon 1.9 eV excitation. As a result, interfacial excitons are formed with holes residing in the valence band of graphene and electrons remaining in the conduction band of $MoS_2$. The subsequent relaxation of the THz photoconductivity comes from the interfacial exciton recombination.

For $MoS_2$/Gr, as shown in **Figure 3**d, a slow-decay component appears with lifetime longer than 1 ns, **Figures 3**e and **3**f present the fitted lifetimes and amplitude weights**,** respectively. Similarly, the fast lifetime $\tau_1$ (1~2 ps) comes from graphene thermal electron cooling. It is worth noting that the relaxation time $\tau_2$ corresponding to the positive photoconductivity component caused by CT is on the order of ~ ns, *i.e.*, the formation of long-lived charge separation state following ultrafast carrier transport in the $MoS_2$/Gr heterostructure. Moreover, **Figure S3c** shows the transient photoconductivity under higher energy photoexcitation (3.2 eV), with a recovery time of over ~ 1.5 ns. Here, we proposed that the long-lived $\sim\tau_2$ could come from the capture of defect states in $MoS_2$ layer, which has been reported in the previous literature.[26] For $MoS_2$/Gr heterostructure, the $MoS_2$ layer is located on the top of the heterostructure that could be oxidized by prolonged exposure to air, leading to the formation of more defects. With high energy photoexcitation, electrons in the $MoS_2$ conduction band are rapidly captured by the defect states, while the valence band holes are directly injected to the graphene layer leads to an increase in THz photoconductivity. The following ns time scale positive photoconductivity is ascribed to the interfacial recombination of graphene holes with $MoS_2$ defect-state electrons. Furthermore, by comparing the fitting weight $A_2$ of Gr/$MoS_2$ (~ 45 %) and $MoS_2$/Gr (~ 15%) as shown in **Figure** 3c and **3**f, we notice that the relative proportion of the interfacial carrier recombination is three-fold times less in $MoS_2$/Gr, *i.e.*, the number of transferred holes is minor and the vast majority of holes are still retained in the valence band of $MoS_2$ layer, this is due to the hindering effect of the substrate electric field in the $MoS_2$/Gr heterostructure. Unlike thermal electrons crossing the barrier, a small number of holes are still allowed to be directly transferred in the presence of the substrate electric field. **Figures S3a and S3b** present a detailed comparison of the photoconductivity for the Gr/$MoS_2$ and $MoS_2$/Gr heterostructures, respectively, under above A-exciton excitation with various pump fluences.

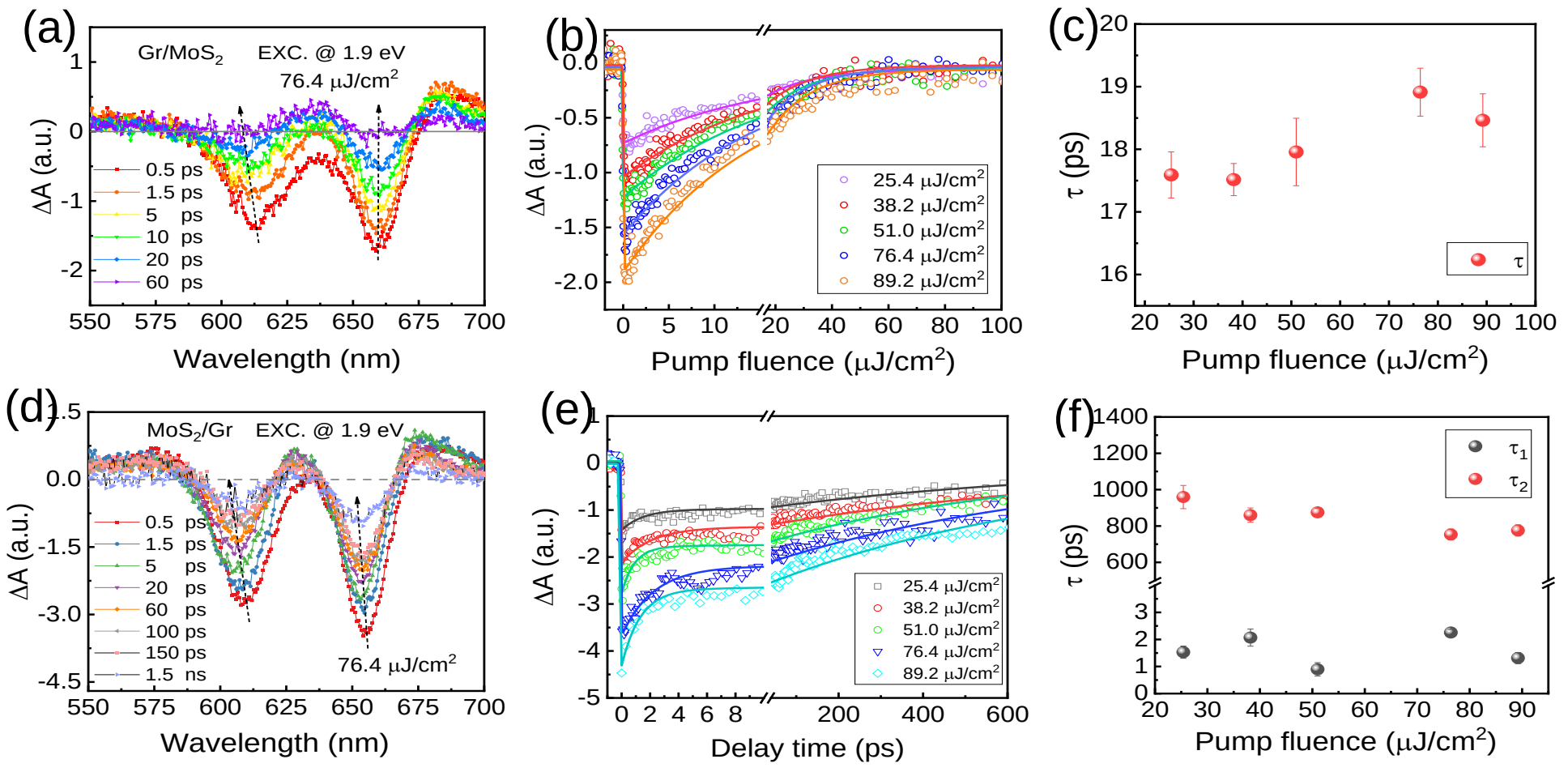

Figure 4. TAS of Gr-$MoS_2$ heterostructures at 1.9 eV (650 nm) excitation. (a and f) Representative TAS of the Gr/$MoS_2$ and $MoS_2$/Gr heterostructures probed at different delay times, respectively, and the pump fluence was fixed at 76.4 $\mu$J cm$^{-2}$. (b and e) The attenuation dynamics of A-exciton state

of the Gr/$MoS_2$ and the $MoS_2$/Gr under several pump fluences, where the solid lines are the fitted curves of the exponential decay function. (e and f) The recovery lifetimes of A-exciton states in Gr/$MoS_2$ and $MoS_2$/Gr heterostructures are fitted by single and double exponential functions, respectively.

TAS with above A-exciton excitation is also conducted to probe the A- and B-exciton dynamics of $MoS_2$ layer in the heterostructures. **Figure 4**a and **4d** show the TA spectra of Gr/$MoS_2$ and $MoS_2$/Gr heterostructures collected at different delay times under pump fluence of 76.4 $\mu$J $cm^{-2}$ and photoenergy of 1.9 eV, respectively. The pronounced photobleaching signals around the A- (660 nm) and B-exciton (610 nm) peaks are clearly seen, which arises from the conduction band of $MoS_2$ being populated with electrons after photoexcitation. The simultaneous appearance of photobleaching for both A- and B-excitons further suggests that on-resonance excitation leads to the transfer of holes，rather than electrons， from $MoS_2$ to graphene layers in both types of heterostructures. In agreement with the THz observation in **Figure 3**a, the excitonic decay process in Gr/$MoS_2$ is completed within dozens of ps timescale, in contrast, the excitonic dynamics in the $MoS_2$/Gr heterostructure is found to occur on timescale of ~ ns. In particular, we focus on the dynamical processes near 660 nm, corresponding to the A exciton transition in ML $MoS_2$, as shown in **Figure 4**b. It is clear that the relaxation for the Gr/$MoS_2$ heterostructure can be well reproduced with single-exponential function, with the fitting data shown in **Figure 4**c. Obviously the fitting lifetime is about 18 ps for different pump fluences, this is very close to the hole lifetime of graphene obtained from the THz spectrum in **Figure 3**b. Therefore, it is conclusive that the direct transfer of $MoS_2$ holes to the graphene layer occurs in the Gr/$MoS_2$ heterostructure with resonant photoexcitation, as shown in **Figure 5**e, and subsequently the electron in $MoS_2$ can combine with the hole in the graphene layer to form an interfacial exciton with typical lifetime of 18 ps, this shows the exactly same way as that under sub-A-exciton photoexcitation. **Figure S4a** in the SI shows the A-exciton photobleaching recovery signals of the Gr/$MoS_2$ at both sub-A-exciton and resonant excitation, it is seen that both of which decay at the identical timescale. However, for the $MoS_2$/Gr heterostructure, as shown in **Figure 4**e, the transient kinetic process conforms to the biexponential characteristics, and the fitting results are presented in **Figure 4**f. The fast component with lifetime of 1~2 ps ($\tau_1$) results from the trapping of excited electrons by the defect states of the $MoS_2$ layer. It is noted that the same process was also observed in ML $MoS_2$ as shown in **Figure S4b in the SI.** In addition, a long-lived process (~ $\tau_2$) appears in the $MoS_2$/Gr heterostructure even slower than that in ML $MoS_2$, which is attributed to the interfacial recombination of the electrons of $MoS_2$ defect states with the holes in graphene. Clearly the observation from TAS is consistent with the THz data in **Figures 3**d and 3e. In short, direct hole transfer from the $MoS_2$ valence band to the graphene layer occurs in both the Gr/$MoS_2$ and $MoS_2$/Gr heterostructures with the photoexcitation of above A-exciton of $MoS_2$. Furthermore, we have demonstrated that the formation of interfacial excitons in the heterostructures following CT, and the interfacial excitons show a typical lifetime of 18 ps in the Gr/$MoS_2$ heterostructure. However, the charge recombination time increases significantly in the $MoS_2$/Gr heterostructure due to the trapping effect of defect states in the top layer of $MoS_2$. **Figures 5**e and **5**f schematically show the charge separation and interfacial exciton for the Gr/$MoS_2$ and $MoS_2$/Gr heterostructures with resonant excitations, respectively.

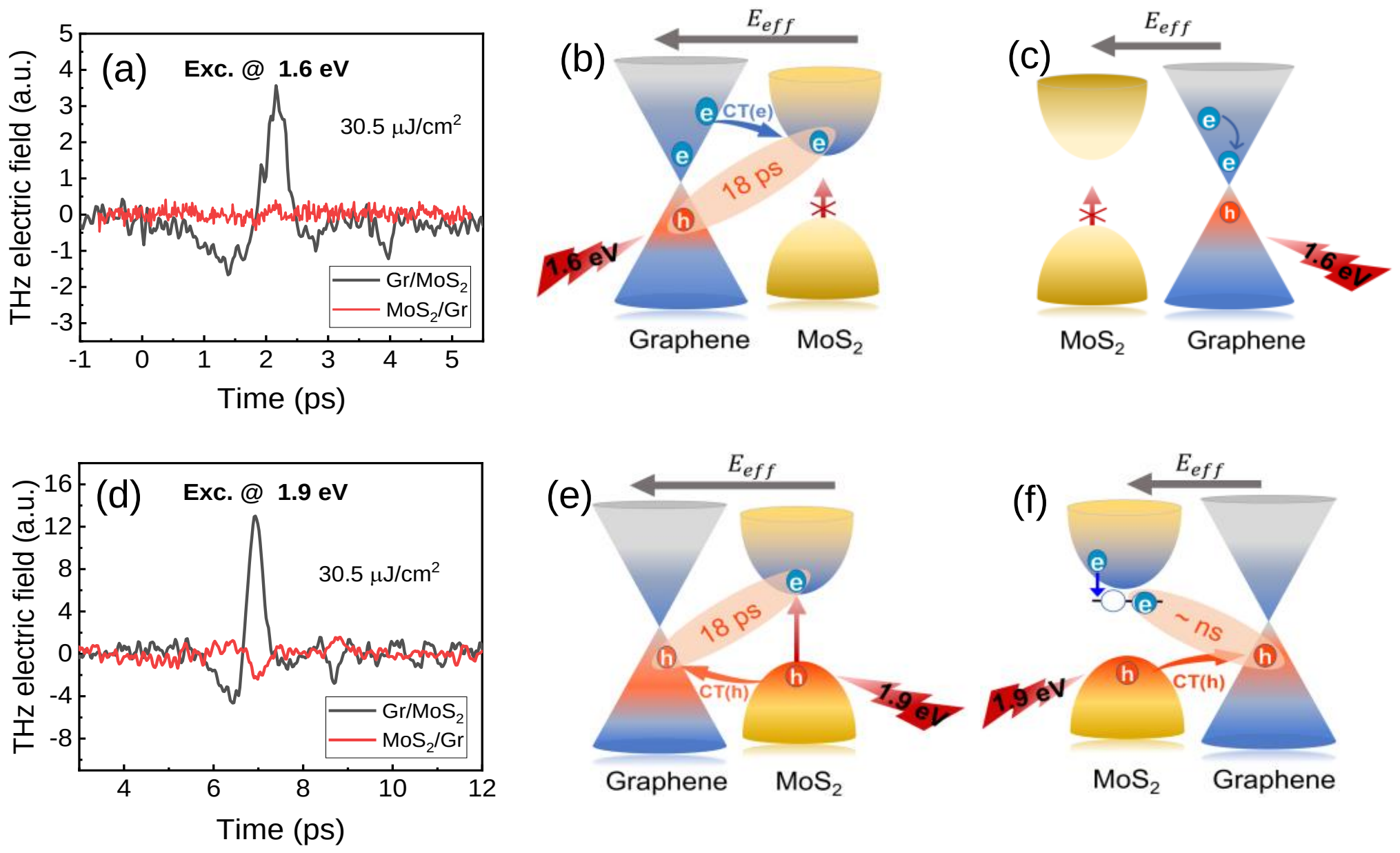

Figure 5. TES and CT maps for the Gr-$MoS_2$ heterostructures with below and above A-exciton

excitations. (a-c) The Emission THz electric field waveforms and CT diagrams of the Gr/$MoS_2$ and $MoS_2$/Gr heterostructures, respectively, with below A-exciton photoexcitation of ~1.6 eV. (d-f) The Emission THz electric field waveforms and CT diagrams of heterostructures of Gr/$MoS_2$ and $MoS_2$/Gr, respectively, with above A-exciton photoexcitation of ~1.9 eV.

As a further confirmation of our point, we conducted TES measurements on the two heterostructures. The TES allow the detection of transient interfacial charge separation processes along the stacking direction of the vertical Gr-$MoS_2$ heterostructures under photoexcitation. In the experiment, we adjusted the excitation pulses to irradiate the Gr-$MoS_2$ heterostructures with incident angles θ=0° and 45°, respectively, and no THz emission signal was detected under normal incident, *i.e.* θ=0°, proving that the far-field THz waves all come from the out-of-plane charge separation in the heterostructures. All subsequent data were measured under 45° incidence, as shown in **Figure 1**b. Similarly, we tune the pump pulse energy to 1.6 eV, below the A-exciton of $MoS_2$, **Figure 5**a shows a typical THz radiation under pump fluence of 30.5 μJ $cm^{-2}$. It is clear that the strong THz emission signal in the Gr/$MoS_2$ heterostructure provides direct and strong evidence for the interfacial CT scenario. As illustrated in **Figure 5**b, the hot electrons from the graphene layer are injected into the $MoS_2$ layer following optical excitation, generating interfacial ultrafast photocurrent, as a result, THz radiation is observed in far field. However, the absence of the THz emission signal in the $MoS_2$/Gr additionally proves that the interfacial charge separation is blocked. Due to the obstruction of the electric field by substrate, as shown in **Figure 5**c, the hot electrons in the graphene layer remain in the layer for cooling without interfacial transport. **Figure S5a in SI** presents the comparison of THz radiation with pump pulse incident from front (film side) and back side (substrate side), the inverse polarity of the THz radiation further suggests that the observed THz signal arises from dynamical CT across the interface between graphene and $MoS_2$. Additionally, it is also seen that the emitted THz pulse reverses its polarity by changing the incident angle θ from $+45^0$ to $-45^0$ (**Figure S5b**), this further demonstrates that the detected THz radiation is rooted from the out-of-plane transient photocurrent, that is contributed by the projection of photocurrent along pump propagation direction *via* CT.

**Figure 5**d records the THz radiation under resonance excitation of A-exciton, with a pump fluence of 30.5 $\mu$J cm$^{-2}$. Notably, after switching the stacking sequence, the two types of heterostructures emit THz pulses with the opposite polarity, that is, the photocurrent due to CT at the interface depends on the stacking sequence, indicating that the CT direction is the same between the two types after photoexcitation. **Figures 5**e and **5**f schematically show the interfacial charge distribution of the two heterostructures under such excitation, where direct hole transfer from the $MoS_2$ valence band to the graphene valence band occurs in both the Gr/$MoS_2$ and $MoS_2$/Gr heterostructures. The polarity of the THz pulse emitted from the two heterostructures also changes as the pump pulse is incident from the backward side, as shown in **Figure S5c**. In addition, there is no detectable THz radiation signal from ML $MoS_2$ at various excitation angles (**Figure S5d**), which rules out a possible optical migration effect from the $MoS_2$ layer. Remarkably, by comparing **Figures 5**a and **5**d, we observe that the radiated THz intensity in the Gr/$MoS_2$ heterostructure with 1.9 eV photoexcitation is 3 times higher than that with 1.6 eV, indicating a higher CT efficiency. This is due to the bidirectional superposition process: electrons injection from graphene into the $MoS_2$ layer as well as the transfer of $MoS_2$ holes to the graphene layer under the strong optical absorption of $MoS_2$. Conversely, the inverted stacked $MoS_2$/Gr heterostructure exhibits a weaker THz radiation signal, indicating a lower interfacial charge separation efficiency and supporting the photoconductivity results observed in **Figure 3**d, which can be interpreted as a blocking effect of the surface-oriented substrate electric field in $MoS_2$/Gr, resulting in fewer holes being injected into the graphene by the $MoS_2$ layer.

## IV. CONCLUSIONs

To summarize, combined with the observation of the interfacial charge dynamics of Gr-$MoS_2$ vertical heterostructures with different photoexcitation energies and stacking sequences by means of the diverse spectroscopy of TRTS, TAS and TES, we conclude that the graphene layer thermal electrons in the Gr/$MoS_2$ heterostructure are effectively injected into the $MoS_2$ layer *via* excitation below the A-exciton, while the interfacial thermal carrier transport in the $MoS_2$/Gr heterostructure is blocked by the electric field of the substrate. On the other hand, with above A-exciton excitation, the injection of $MoS_2$ valence band holes into the graphene layer occurs in both Gr/$MoS_2$ and $MoS_2$/Gr heterostructures, and the resulting interfacial photocurrents are generated in the two heterostructures with the opposite directions, this is evidenced by THz emission with the opposite polarity. Furthermore, we show that once CT occurs in the Gr-$MoS_2$ heterostructures, the subsequent relaxation of the interfacial charges takes place on a lifetime scale of 18 ps to 1 ns, most likely depending on the density of defect states in $MoS_2$ layer. These measurements of the complex interfacial charge dynamics of graphene-based heterojunctions provide new light for regulating the directionality and efficiency of the photogating field in vdW heterostructures. It has important scientific implications for the design and performance optimization of optoelectronic devices in 2D materials.

## V. REFERENCEs

(1) Castro Neto, A. H.; Guinea, F.; Peres, N. M. R.; Novoselov, K. S.; Geim A. K. The Electronic Properties of Graphene. *Rev. Mod. Phys.* **2009**, *81*, 109-162.

(2) Bolotin, K. I.; Sikes, K. J.; Jiang, Z.; Klima, M.; Fudenberg, G.; Hone, J.; Kim, P.; Stormer, H. L. Ultrahigh Electron Mobility in Suspended Graphene. *Solid State Communications*. **2008**, *146*, 351-355.

(3) Yang, H.; Shen, C.; Tian, Y.; Bao, L.; Chen, P.; Yang, R.; Yang, T.; Li, J.; Gu, C.; Gao, H.-J. High-Quality Graphene Grown on Polycrystalline PtRh20 Alloy Foils by Low Pressure Chemical Vapor Deposition and Its Electrical Transport Properties. *Appl. Phys. Lett*. **2016**, 108, 063102.

(4) Nair, R. R.; Blake, P.; Grigorenko, A. N.; Novoselov, K. S.; Booth, T. J.; Stauber, T.; Peres, N. M. R.; Geim, A. K. Fine Structure Constant Defines Visual Transparency of Graphene. *Science*. **2008**, *320*, 1308.

(5) Xiao, D.; Liu, G. B.; Feng, W.; Xu, X.; Yao, W. Coupled Spin and Valley Physics in Monolayers of $MoS_2$ and Other Group-VI Dichalcogenides. *Phys. Rev. Lett.* **2012**, *108*, 196802.

(6) Zeng, H.; Dai, J.; Yao, W.; Xiao, D.; Cui, X. Valley Polarization in $MoS_2$ Monolayers by Optical Pumping. *Nat. Nanotechnol.* **2012**, *7*, 490−493.

(7) Radisavljevic, B.; Radenovic, A.; Brivio, J.; Giacometti, V.;Kis, A. Single-Layer $MoS_2$ Transistors. *Nature Nanotechnol.* **2011**, *6*, 147–150.

(8) Liu, K. K.; Zhang, W. J.; Lee, Y. H.; Lin, Y. C.; Chang, M. T.; Su, C.; Chang, C. S.; Li, H.; Shi, Y. M.; Zhang, H.; Lai, C. S.; Li, L. J.; Growth of Large-Area and Highly Crystalline $MoS_2$ Thin Layers on Insulating Substrates. *Nano Lett*. **2012**, *12*, 1538–1544.

(9) Mak, K. F.; Lee, C.; Hone, J.; Shan, J; Heinz, T. F. Atomically Thin $MoS_2$: A New Direct-Gap Semiconductor. *Phys. Rev. lett.* **2010**, *105*, 136805–136808.

(10) Kumar, N.; Najmaei, S.; Cui, Q.; Ceballos, F.; Ajayan, P. M.; Lou, J.; Zhao, H. Second Harmonic Microscopy of Monolayer $MoS_2$. *Phys. Rev. B: Condens. Matter Mater. Phys.* **2013**, *87*,161403.

(11) Malard, L. M.; Alencar, T. V.; Barboza, A. P. M.; Mak, K. F.; de Paula, A. M. Observation of Intense Second Harmonic Generation from $MoS_2$ Atomic Crystals. *Phys. Rev. B: Condens. Matter Mater. Phys.* **2013**, *87*, 201401.

(12) Li, Y.; Rao, Y.; Mak, K. F.; You, Y.; Wang, S.; Dean, C. R.; Heinz, T. F. Probing Symmetry Properties of Few-Layer $MoS_2$ and h-BN by Optical Second-Harmonic Generation. *Nano Lett.* **2013**, *13*, 3329−3333.

(13) Yuan, L.; Chung, T.-F.; Kuc, A.; Wan, Y.; Xu, Y.; Chen, Y. P.; Heine, T.; Huang, L. Photocarrier Generation from Interlayer Charge Transfer Transitions in $WS_2$-Graphene Heterostructures. *Sci. Adv.* **2018**, *4*, No. e1700324.

(14) Froehlicher, G.; Lorchat, E.; Berciaud, S. Charge Versus Energy Transfer in Atomically Thin Graphene-Transition Metal Dichalcogenide van der Waals Heterostructures. *Phys. Rev. X.* **2018**, *8*, 011007.

(15) He, J.; Kumar, N.; Bellus, M. Z.; Chiu, H.-Y.; He, D.; Wang, Y.; Zhao, H. Electron Transfer and Coupling in Graphene-Tungsten Disulfide van der Waals Heterostructures. *Nat. Commun.* **2014**, *5*, 5622.

(16) Yang, B.; Tu, M.-F.; Kim, J.; Wu, Y.; Wang, H.; Alicea, J.; Wu, R.; Bockrath, M.; Shi, J. Tunable Spin-Orbit Coupling and SymmetryProtected Edge States in Graphene/$WS_2$. *2D*

*Mater.* **2016**, *3*, 031012.

(17) Wang, Z.; Ki, D. K.; Chen, H.; Berger, H.; MacDonald, A. H.; Morpurgo, A. F. Morpurgo, Strong Interface-Induced Spin-Orbit Interaction in Graphene on $WS_2$. *Nat. Commun.* **2015**, *6*, 8339.

(18) Shim, J.; Kang, D.-H.; Kim, Y.; Kum, H.; Kong, W.; Bae, S.-H.; Almansouri, I.; Lee, K.; Park, J.-H.; Kim, J. Recent Progress in van der Waals (vdW) Heterojunction-Based Electronic and Optoelectronic Devices. *Carbon*. **2018**, *133*, 78−89.

(19) Zhang, W.; Chuu, C.-P.; Huang, J.-K.; Chen, C.-H.; Tsai, M.-L.; Chang, Y.-H.; Liang, C.-T.; Chen, Y.-Z.; Chueh, Y.-L.; He, J.-H.; Chou, M.-Y.; Li, L.-J. Ultrahigh-Gain Photodetectors Based on Atomically Thin Graphene-$MoS_2$ Heterostructures. *Sci. Rep.* **2014**, *4*, 3826.

(20) Georgiou, T.; Jalil, R.; Belle, B. D.; Britnell, L.; Gorbachev, R. V.; Morozov, S. V.; Kim, Y.-J.; Gholinia, A.; Haigh, S. J.; Makarovsky, O.; Eaves, L.; Ponomarenko, L. A.; Geim, A. K.; Novoselov, K. S.; Mishchenko, A. Vertical Field-Effect Transistor Based on Graphene-$WS_2$ Heterostructures for Flexible and Transparent Electronics. *Nat. Nanotechnol.* **2013**, *8*, 100−103.

(21) Rodriguez-Nieva, J. F.; Dresselhaus, M. S.; Levitov, L. S. Thermionic Emission and Negative dI/dV in Photoactive Graphene Heterostructures. *Nano Lett.* **2015**, *15*, 1451−1456.

(22) Bai, L.; Wang, X.; Tang, S.; Kang, Y.; Wang, J.; Yu, Y.; Zhou, Z. K.; Ma, C.; Zhang, X.; Jiang, J.; Chu, P. K.; Yu, X.-F. Black Phosphorus/Platinum Heterostructure: A Highly Efficient Photocatalyst for Solar-Driven Chemical Reactions. *Adv. Mater.* **2018**, *30*, 1803641.

(23) Guo, J.; Zhang, Y.; Shi, L.; Zhu, Y.; Mideksa, M. F.; Hou, K.; Zhao, W.; Wang, D.; Zhao, M.; Zhang, X.; Lv, J.; Zhang, J.; Wang, X.; Tang, Z. Boosting Hot Electrons in Hetero-Superstructures for Plasmon-Enhanced Catalysis. *J. Am. Chem. Soc.* **2017**, *139*, 17964−17972.

(24) Massicotte, M.; Schmidt, P.; Vialla, F.; Watanabe, K.; Taniguchi, T.; Tielrooij, K. J.; Koppens, F. H. L. Photo-thermionic Effect in Vertical Graphene Heterostructures. *Nat. Commun.* **2016**, *7*, 12174.

(25) Zou, Y. Q.; Ma, Q.-S.; Zhang, Z. Y.; Pu, R. H.; Zhang, W. J.; Sou, P.; Sun, K. W.; Chen, J. M.; Li, D.; Ma, H.; Lin, X.; Leng, Y. X.; Liu, W. M.; Du, J.; Ma, G. H. Observation of Ultrafast Interfacial Exciton Formation and Relaxation in Graphene/$MoS_2$ Heterostructure. *J. Phys.*

*Chem. Lett.* **2022**, *13*, 5123−5130.

(26) Fu, S.; du Fossé, I.; Jia, X.; Xu, J. Y.; Yu, X. Q.; Zhang, H.; Zheng, W. H.; Krasel, S.; Chen, Z. P.; Wang, Z. M.; Tielrooij, K.-J.; Bonn, M., Houtepen, A. J.; Wang, H. I. Long-lived Charge Separation Following Pump Wavelength Dependent Ultrafast Charge Transfer in Graphene/$WS_2$ Heterostructures. *Sci. Adv.* **2021**, *7*, No. eabd9061.

(27) Chen, Y.; Li, Y.; Zhao, Y.; Zhou, H.; Zhu, H. Highly Efficient Hot Electron Harvesting from Graphene Before Electron-Hole Thermalization. *Sci. Adv.* **2019**, *5*, No. eaax9958.

(28) Trovatello, C.; Piccinini, G.; Forti, S.; Fabbri, F.; Rossi, A.; De Silvestri, S.; Coletti, C.; Cerullo, G.; Dal Conte, S. Ultrafast Hot Carrier Transfer in $WS_2$/Graphene Large Area Heterostructures. *Npj 2D Mater. Appl.* **2022**, *6*, 24.

(29) Lin, Y.; Ma, Q.; Shen, P.-C.; Ilyas, B.; Bie, Y.; Liao, A.; Ergeçen, E.; Han, B.; Mao, N.; Zhang, X.; Ji, X.; Zhang, Y. H.; Yin, J. H.; Huang, S. X.; Dresselhaus, M.; Gedik, N.; Jarillo-Herrero, P.; Ling, X.; Kong, J.; Palacios, T. Asymmetric Hot-Carrier Thermalization and Broadband Photoresponse in Graphene-2D Semiconductor Lateral Heterojunctions. *Sci. Adv.* **2019**, *5*, No. eaav1493.

(30) Aeschlimann, S.; Rossi, A.; Chavez-Cervantes, M.; Krause, R.; Arnoldi, B.; Stadtmuller, B.; Aeschlimann, M.; Forti, S.; Fabbri, F.; Coletti, C.; Gierz, I. Direct Evidence for Efficient Ultrafast Charge Separation in Epitaxial $WS_2$/Graphene Heterostructures. *Sci. Adv.* **2020**, *6*, No. eaay0761.

(31) Luo, D.; Tang, J.; Shen, X.; Ji, F.; Yang, J.; Weathersby, S.; Kozina, M. E.; Chen, Z.; Xiao, J.; Ye, Y. S.; Cao, T.; Zhang, G.Y.; Wang, X. J.; Lindenberg, A. M. Twist-Angle-Dependent Ultrafast Charge Transfer in $MoS_2$-Graphene van Der Waals Heterostructures. *Nano Lett.* **2021**, *2*, 8051−8057.

(32) Ferrante, C.; Di Battista, G.; López, L. E. P.; Batignani, G.; Lorchat, E.; Virga, A.; Berciaud, S.; Scopigno, T. Picosecond Energy Transfer in a Transition Metal Dichalcogenide-Graphene Heterostructure Revealed by Transient Raman Spectroscopy. *Proc. Natl. Acad. Sci. U. S. A.* **2022**, *119*, No. e2119726119.

(33) Liu, H.; Wang, J.; Liu, Y.; Wang, Y.; Xu, L.; Huang, L.; Liu, D.; Luo, J. Visualizing Ultrafast Defect-Controlled Interlayer Electronphonon Coupling in van Der Waals Heterostructures. *Adv.*

*Mater.* **2022**, *34*, 2106955.

(34) Krause, R.; Aeschlimann, S.; Chávez-Cervantes, M.; PereaCausin, R.; Brem, S.; Malic, E.; Forti, S.; Fabbri, F.; Coletti, C.; Gierz, I. Microscopic Understanding of Ultrafast Charge Transfer in van Der Waals Heterostructures. *Phys. Rev. Lett.* **2021**, *127*, 276401.

(35) Liu, Y. X.; Zhang, J.; Meng, S.; Yam, C. Y.; Frauenheim, T. Electric Field Tunable Ultrafast Interlayer Charge Transfer in Graphene/$WS_2$ Heterostructure. *Nano Lett.* **2021**, *21*, 4403−4409.

(36) Fu, S.; Jia, X. Y.; Hassan, A. S.; Zhang, H.; Zheng, W. H.; Gao, L.; Di Virgilio, L.; Krasel, S.; Beljonne, D.; Tielrooij, K.-J.; Bonn, M.; Wang, H. I. Reversible Electrical Control of Interfacial Charge FlowAcross van der Waals Interfaces. *Nano Lett.* **2023**, *23*, 1850-1857.

(37) Dhakal, K. P.; Duong, D. L.; Lee, J.; Nam, H.; Kim, M.; Kan, M.; Lee, Y. H.; Kim, J. Confocal Absorption Spectral Imaging of $MoS_2$: Optical Transitions Depending on the Atomic Thickness of Intrinsic and Chemically Doped $MoS_2$. *Nanoscale.* **2014**, *6*, 13028.

(38) Wang, L.; Wang, Z.; Wang, H. Y.; Grinblat, G.; Huang, Y. L.; Wang, D.; Ye, X. H.; Li, X. B.; Bao, Q. L.; Wee, A. S.; Maier, S. A.; Chen, Q. D.; Zhong, M. L.; Qiu, C. W.; Sun, H.-B. Slow Cooling and Efficient Extraction of C-exciton Hot Carriers in $MoS_2$ Monolayer. *Nat. Commun.* **2017**, *8*, 13906.

(39) Lu, A. Y.; Wei, S. Y.; Wu, C. Y.; Hernandez, Y.; Chen, T. Y.; Liu, T. H.; Pao, C. W.; Chen, F. R.; Li, L. J.; Juang, Z. Y. Decoupling of CVD Graphene by Controlled Oxidation of Recrystallized Cu. *RSC Adv.* **2012**, *2*, 3008−3013.

(40) Ghopry, S. A.; Alamri, M. A.; Goul, R.; Sakidja, R.; Wu, J. Z. Extraordinary Sensitivity of Surface-Enhanced Raman Spectroscopy of Molecules on $MoS_2$ ($WS_2$) Nanodomes/Graphene van der Waals Heterostructure Substrates. *Adv. Opt. Mater.* **2019**, *7*, 1801249.

(41) Xing, X., Zhao, L. T.; Zhang, W. J.; Wang, Z.; Su, H. M.; Chen, H. Y.; Ma, G. H.; Dai, J. F.; Zhang, W. J.; Influence of a Substrate on Ultrafast Interfacial Charge Transfer and Dynamical Interlayer Excitons in Monolayer $WSe_2$/Graphene Heterostructures. *Nanoscale.* **2020**, 12, 2498-2506.

(42) Ma, Q.-S.; Zhang, W. J.; Wang, C. L.; Pu, R. H.; Ju, C.-W.; Lin, X.; Zhang, Z. Y.; Liu, W. M.; Li, R. X. Hot Carrier Transfer in a Graphene/$PtSe_2$ Heterostructure Tuned by a Substrate-Introduced Effective Electric Field. J. Phys. Chem. C 2021, 125, 9296−9302.

(43) Srivastava, Y. K.; Chaturvedi, A.; Manjappa, M.; Kumar, A.; Dayal, G.; Kloc, C.; Singh, R. $MoS_2$ for Ultrafast All-Optical Switching and Modulation of THz Fano Metaphotonic Devices. *Adv. Opt. Mater.* **2017**, *5*, 1700762.

(44) Ma, E. Y.; Guzelturk, B.; Li, G. Q.; Cao, L. Y.; Shen, Z. X.; Lindenberg, A. M.; Heinz, T. F. Recording Interfacial Currents on the Subnanometer Length and Femtosecond Time Scale by Terahertz Emission. *Sci. Adv.* **2019**, *5*, No. eaau0073.

(45) Yang, J.; Jiang, S. L.; Xie, J. F.; Jiang H. C.; Xu, S. J.; Zhang, K.; Shi, Y. P.; Zhang, Y. F.; Zeng, Z.; Fang, G. Y. Identifying the Intermediate Free-Carrier Dynamics Across the Charge Separation in Monolayer $MoS_2/ReSe_2$ Heterostructures. *ACS Nano.* **2021**, *15*, 16760−16768.

## VI. SUPPORTING INFORMATIONs

### Section S1: Optical characterization and illustration of effective electric field by substrate.

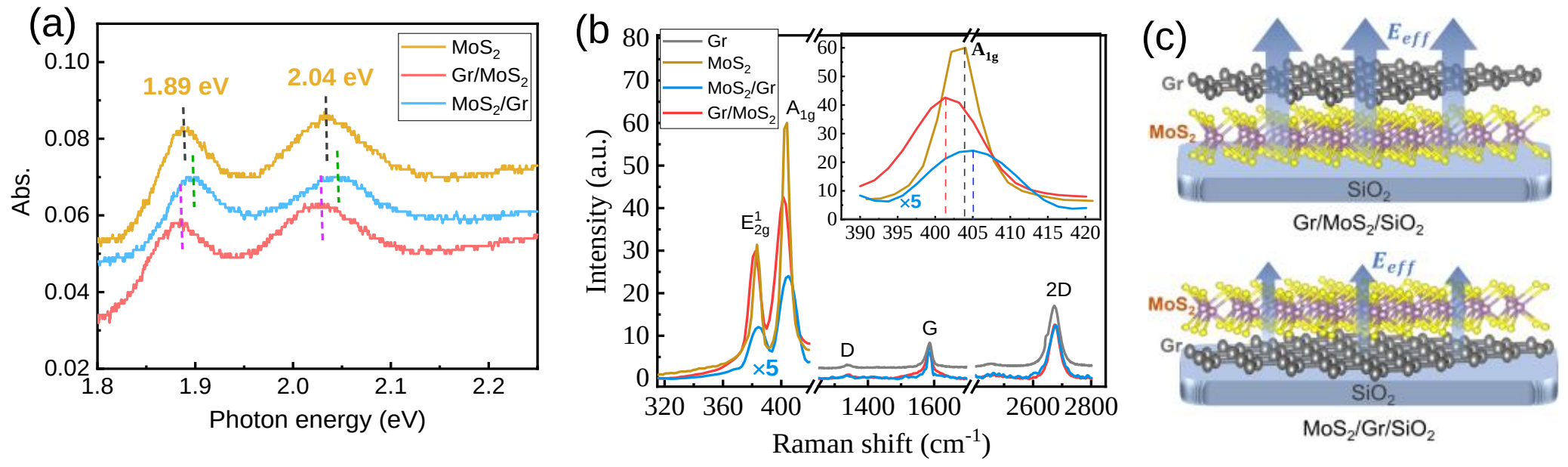

Figure S1. (a) UV-vis absorption spectra of monolayer (ML) $MoS_2$ and Gr-$MoS_2$ heterostructures. (b) Raman spectra of ML Gr, ML $MoS_2$ and Gr-$MoS_2$ heterostructures. (c) Diagram of the effective electric field introduced by $SiO_2$ substrate.

The individual ML $MoS_2$ shows two optical resonances at 1.89 and 2.04 eV in the absorption spectrum of **Figure S1a**, corresponding to the A- and B-exciton transitions of $MoS_2$,[S1,S2] indicating the high-quality monolayer nature of $MoS_2$. In contrast, the exciton peaks of the Gr/$MoS_2$ and $MoS_2$/Gr heterostructures are red-shifted and blue-shifted by about 60 meV, respectively, indicating that the strong interlayer coupling effect of the heterostructures modulates the optical band gap of the $MoS_2$ layer.

**Figure S1b** displays the Raman spectra of samples obtained with 532 nm excitation wavelength, where the Raman intensity of the $MoS_2$/Gr heterostructure has been multiplied by a factor of 5 for better comparison. In Gr-$MoS_2$ heterostructures, except for the characteristic peaks of G-band (~1587 $cm^{-1}$) and 2D-band (~2675 $cm^{-1}$) of graphene layer,[S3,S4] the typical Raman modes of $E_{1g}^{2}$ and $A_{1g}$ of $MoS_2$ layer were observed at ~382 $cm^{-1}$ and ~401 $cm^{-1}$,[S5,S6] respectively. The Fermi level of the graphene layer can be obtained from the G-band characteristic peak, but the G-band position is susceptible to stress, so we can judge the charge doping of each atomic layer in heterostructures based on the $A_{1g}$ peak shift of the $MoS_2$ layer.[S7,S8] In contrast to ML $MoS_2$, the $A_{1g}$ mode of the Gr/$MoS_2$ heterostructure is red-shifted by ~3 $cm^{-1}$, suggesting that additional electrons are injected into the $MoS_2$ layer after exposed to light in Gr/$MoS_2$.[S6,S9] The reverse-stacked $MoS_2$/Gr, in turn, has a slight blue-shift of ~1 $cm^{-1}$, indicating that the $MoS_2$ layer becomes less n-doped (lower electron density) in the heterostructure, which is caused by the transfer process and adsorption of charged substrate impurity effect. We believe that the situation is caused by the surface-directed effective electric field introduced by the $SiO_2$ substrate, which is stronger in the Gr/$MoS_2$ heterostructure, making it easier to separate the photogenerated electron-hole pairs, as shown in **Figure S1c**. The surface-directed electric field can efficiently modulate the interfacial charge transfer across the heterostructures.

**Section S2: Fitting details for transient THz transmission of ML Gr and Gr-MoS2 heterostructures with below A-exciton photoexcitation (1.6 eV).**

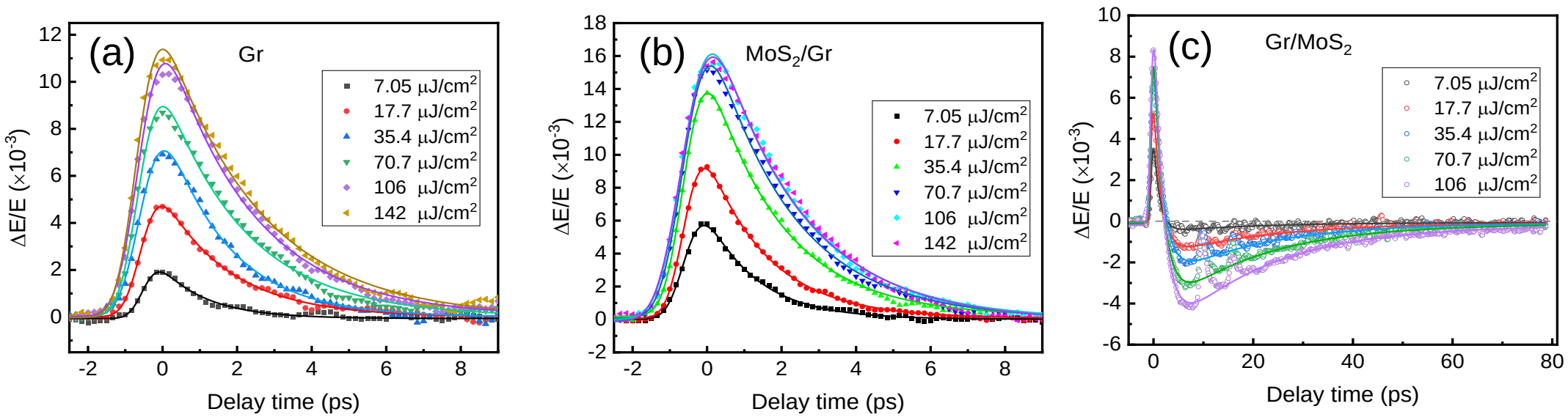

Figure S2. The transient THz response of ML Gr, Gr/$MoS_2$ and $MoS_2$/Gr heterostructures are shown in (a), (b) and (c) respectively, where the solid lines are numerical fitting of the exponential decay function convolved with a 200 fs half-width Gaussian pulse. ML Gr was fitted with a single exponential function, as well as $MoS_2$/Gr, while the Gr/$MoS_2$ heterojunction was fitted with a biexponential decay. The two small peaks located at delay times of 10 and 16 ps in Gr/$MoS_2$ are the secondary reflections from the substrate as well as the terahertz emitter ZnTe, respectively.

**Section S3: The transient THz photoconductivity of Gr/$MoS_2$ and $MoS_2$/Gr heterostructures with above A-exciton photoexcitation.**

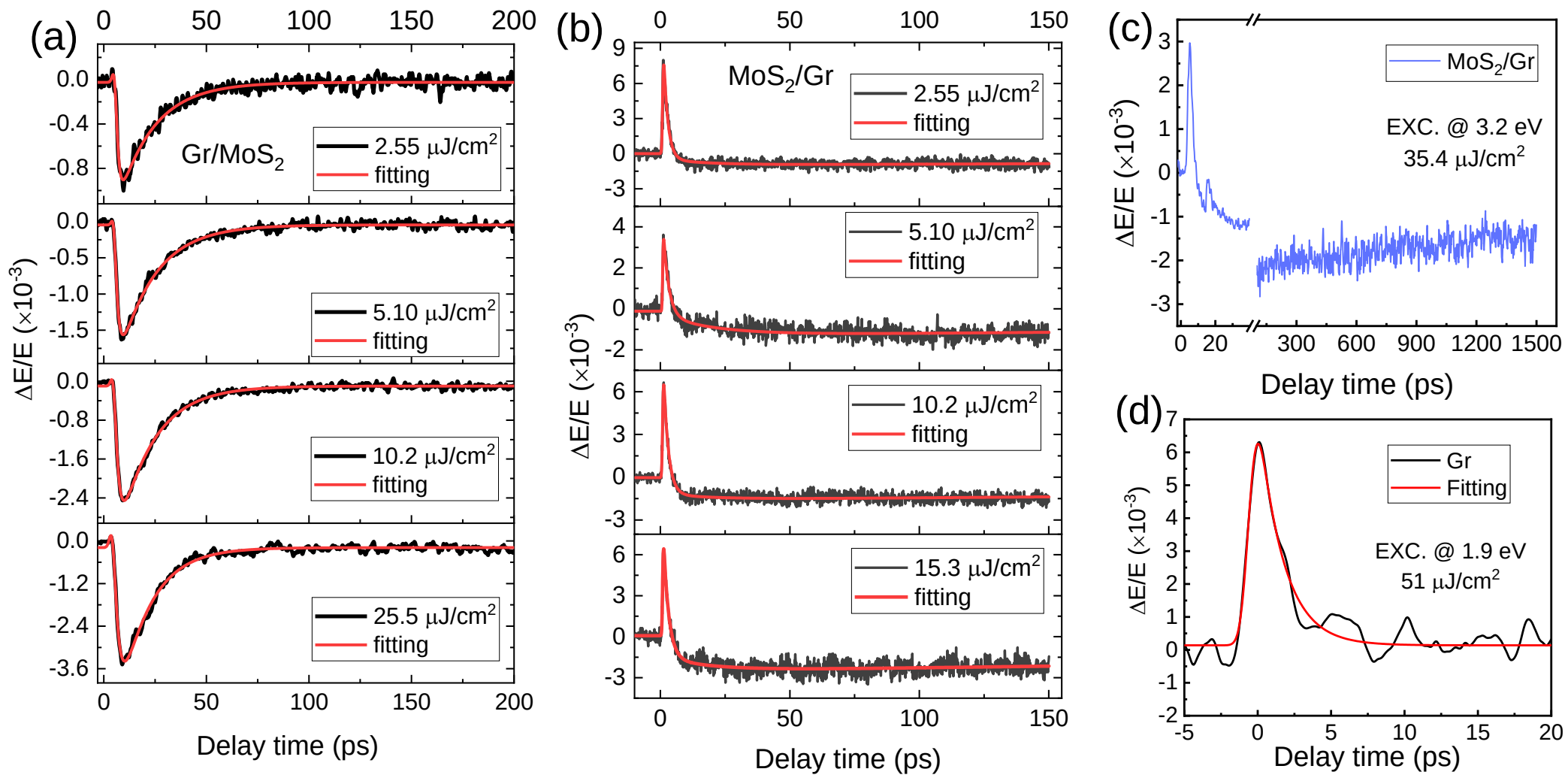

Figure S3. (a) and (b) are the transient THz photoconductivity of Gr/$MoS_2$ and $MoS_2$/Gr heterostructures under 1.9 eV photoexcitation, respectively, with various pump fluences, where the red solid lines are the fitted curves for the biexponential decay function. (c) The THz photoconductivity of the $MoS_2$/Gr heterostructure with 3.2 eV photoexcitation, and the long-lived relaxation process is not fully detected due to the limited measurement range. (d) THz photoconductivity of ML Gr under 1.9 eV photoexcitation, where the red solid line is the fitted curve for the single-exponential decay function.

**Section S4: The photobleaching recovery signal of A-exciton of $MoS_2$ monolayer and Gr-$MoS_2$ heterostructures.**

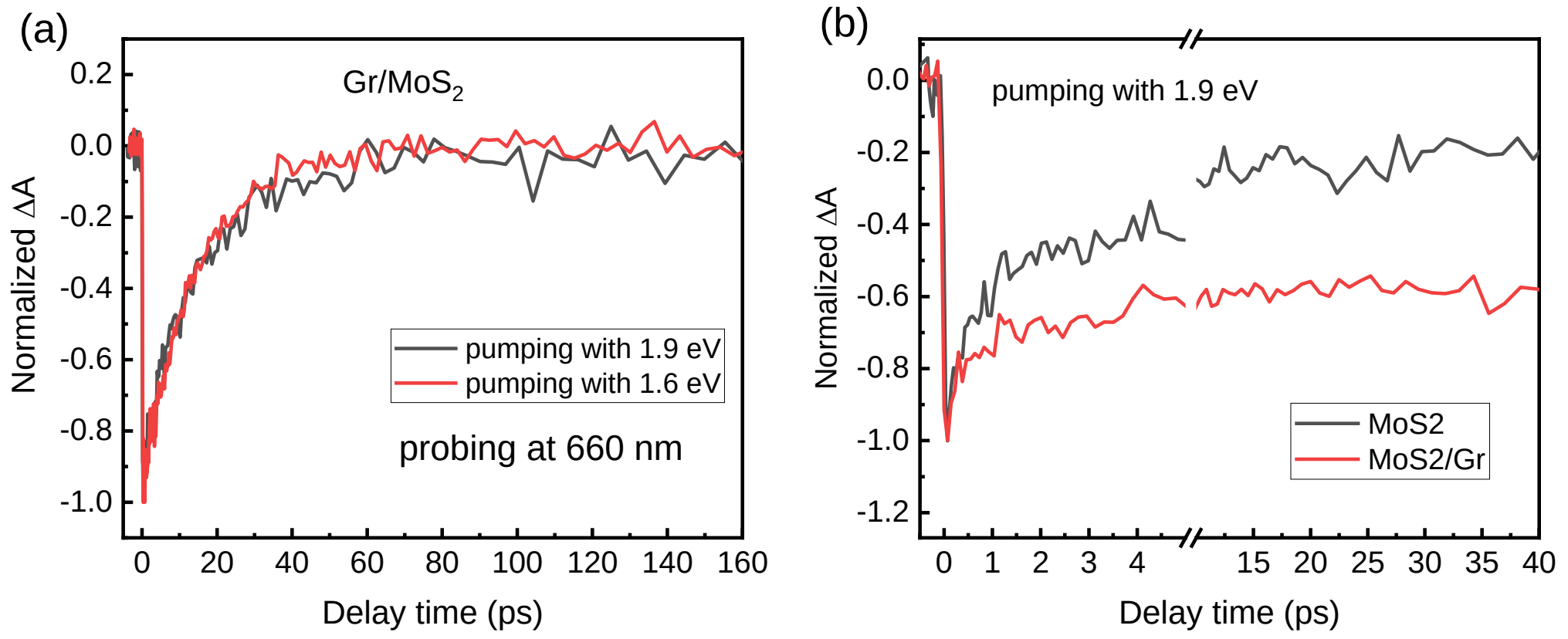

Figure S4. (a) The normalized dynamics of A-exciton of Gr/$MoS_2$ under excitation with photoenergy of 1.6 (780 nm) and 1.9 eV (650 nm). It is clear that the transient dynamics show identical relaxation under excitation of with different photon energy. (b) Normalized transient transmittance of A-exciton of $MoS_2$ monolayer (black) and $MoS_2$/Gr heterostructure (red) under identical pump fluence at wavelength of 650 nm. ML $MoS_2$ and the $MoS_2$/Gr heterostructure decay at the same rapid rate within the initial a few ps, while $MoS_2$/Gr is associated with a longer relaxation lifetime.

**Section S5: THz emission spectra with different incident directions and incident angles.**

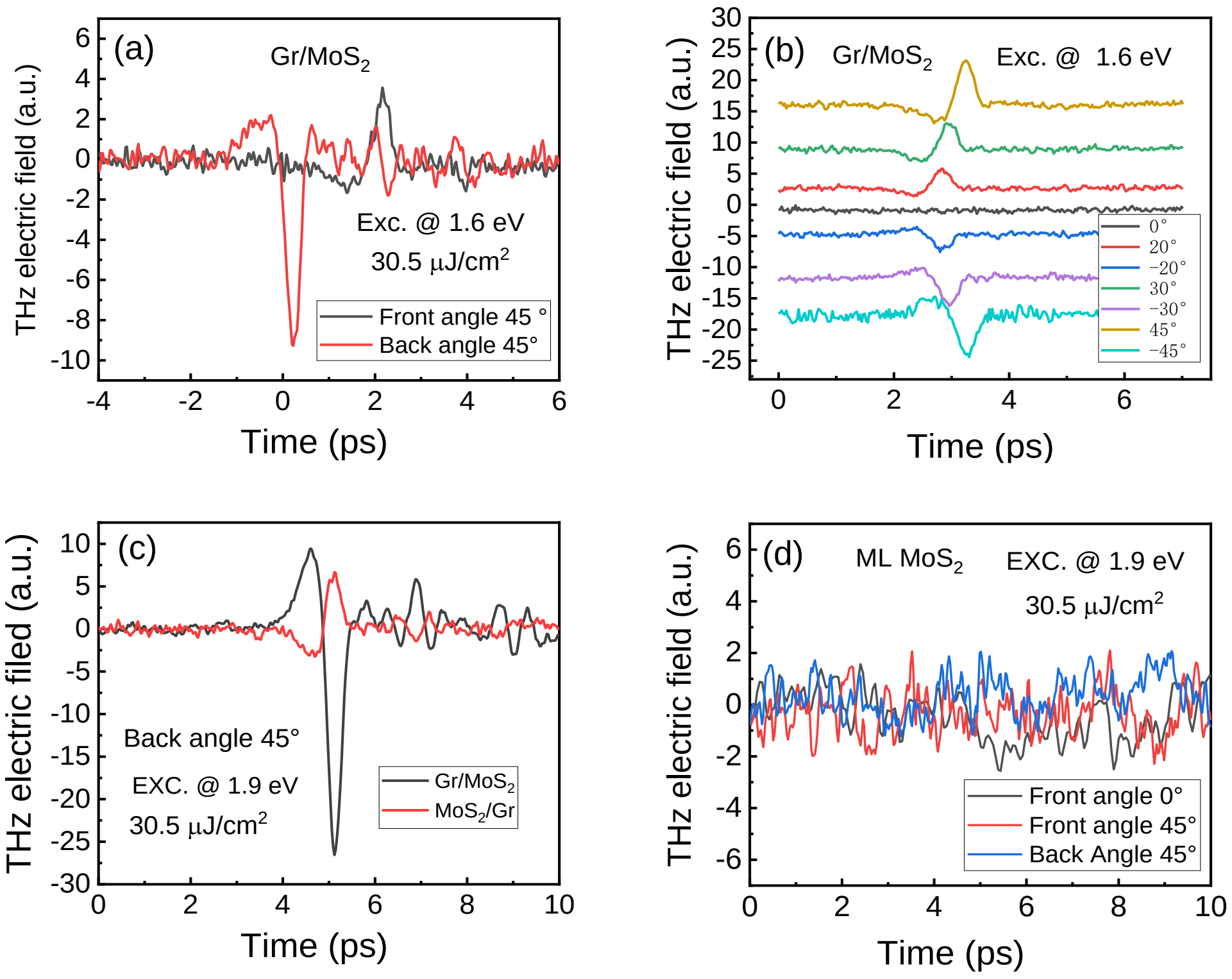

Figure S5. (a) THz radiation of Gr/$MoS_2$ heterostructure under 1.6 eV photoexcitation, the polarity of THz pulse is opposite with that pump pulse is incident from front (film side) and back side (substrate side) at fixed incident angle of 45°, respectively. (b) The emitted THz pulses of the Gr/$MoS_2$ heterostructure at different incident angles, and its polarity is opposite positive and negative incident angles. (c) THz radiation of Gr/$MoS_2$ and $MoS_2$/Gr heterostructures under 1.9 eV photoexcitation, with pump pulse incident at 45° from back side. (d) THz emission spectra of ML $MoS_2$ under 1.9 eV photoexcitation, and no detectable signal with different pump incident directions and incident angles.

## References

S1. Dhakal, K. P.; Duong, D. L.; Lee, J.; Nam, H.; Kim, M.; Kan, M.; Lee, Y. H.; Kim, J. Confocal Absorption Spectral Imaging of MoS2: Optical Transitions Depending on the Atomic Thickness of Intrinsic and Chemically Doped $MoS_2$. *Nanoscale.* **2014**, *6*, 13028.

S2. Wang, L.; Wang, Z.; Wang, H. Y.; Grinblat, G.; Huang, Y. L.; Wang, D.; Ye, X. H.; Li, X. B.; Bao, Q. L.; Wee, A. S.; Maier, S. A.; Chen, Q. D.; Zhong, M. L.; Qiu, C. W.; Sun, H.-B. Slow Cooling and Efficient Extraction of C-exciton Hot Carriers in $MoS_2$ Monolayer. *Nat. Commun.* **2017**, *8*, 13906.

S3. Lu, A. Y.; Wei, S. Y.; Wu, C. Y.; Hernandez, Y.; Chen, T. Y.; Liu, T. H.; Pao, C. W.; Chen, F. R.; Li, L. J.; Juang, Z. Y. Decoupling of CVD Graphene by Controlled Oxidation of Recrystallized Cu. *RSC Adv.* **2012**, *2*, 3008−3013.

S4. Ghopry, S. A.; Alamri, M. A.; Goul, R.; Sakidja, R.; Wu, J. Z. Extraordinary Sensitivity of Surface-Enhanced Raman Spectroscopy of Molecules on $MoS_2$ ($WS_2$) Nanodomes/Graphene van der Waals Heterostructure Substrates. *Adv. Opt. Mater.* **2019**, *7*, 1801249.

S5. Liu, K. K.; Zhang, W. J.; Lee, Y. H.; Lin, Y. C.; Chang, M. T.; Su, C.; Chang, C. S.; Li, H.; Shi, Y. M.; Zhang, H.; Lai, C. S.; Li, L. J.; Growth of Large-Area and Highly Crystalline $MoS_2$ Thin Layers on Insulating Substrates. *Nano Lett*. **2012**, *12*, 1538–1544.

S6. Zhang, W.; Chuu, C.-P.; Huang, J.-K.; Chen, C.-H.; Tsai, M.-L.; Chang, Y.-H.; Liang, C.-T.; Chen, Y.-Z.; Chueh, Y.-L.; He, J.-H.; Chou, M.-Y.; Li, L.-J. Ultrahigh-Gain Photodetectors Based on Atomically Thin Graphene-$MoS_2$ Heterostructures. *Sci. Rep.* **2015**, *4*, 3826.

S7. Chakraborty, B.; Bera, A.; Muthu, D. V. S.; Bhowmick, S.; Waghmare, U. V.; Sood, A. K. Symmetry-Dependent Phonon Renormalization in Monolayer $MoS_2$ Transistor. *Phys. Rev. B*. **2012**, *85*, 161403.

S8. Placidi, M.; Dimitrievska, M.; Izquierdo-Roca, V.; Fontane, X.; Castellanos-Gomez, A.; Perez-Tomas, A.; Mestres, N.; Espindola-Rodriguez, M.; Lopez-Marino, S.; Neuschitzer, M.; Bermudez, V.; Yaremko, A.; Perez-Rodriguez, A. Multiwavelength Excitation Raman Scattering Analysis of Bulk and Two-Dimensional MoS2: Vibrational Properties of Atomically Thin $MoS_2$ Layers. *2D Mater.* ***2015**, 2,* 035006.

S9. Das, A.; Pisana, S.; Chakraborty, B.; Piscanec, S.; Saha, S. K.; Waghmare, U. V.; Novoselov, K. S.; Krishnamurthy, H. R.; Geim, A. K.; Ferrari, A. C.; Sood, A. K. Monitoring Dopants by Raman Scattering in an Electrochemically Top-gated Graphene Transistor. *Nat. Nanotechnol.* **2008**, *3*, 210.